\documentclass{aa}  

\usepackage{graphicx}
\usepackage{txfonts}
\usepackage[switch]{lineno}
\nolinenumbers
\begin{document}

   \title{The benefit of a multi-band high resolution spectroscopic monitoring for studying stellar transients: the NGC~300 OT2008-1 UVES spectrum as a test case. }
   \titlerunning{Low versus high resolution spectroscopy}
    
   \subtitle{}

   \author{Elena Mason
          \inst{1}
          \and
          Steven N. Shore
          \inst{2,1}
          \and 
          Andrea Pastorello 
          \inst{3}
          \and 
          Paolo Di Marcantonio
          \inst{1}
          }

   \institute{INAF-OATS, Via Giambattista Tiepolo 11, 34143 Trieste (TS), Italia\\
              \email{elena.mason@inaf.it}
         \and
             Dipartimento di Fisica “Enrico Fermi”, Università di Pisa, Largo B. Pontecorvo 3, 56127 Pisa (PI), Italia
        \and 
            INAF-OAPD, Vicolo dell’Osservatorio 5, 35122 Padova (PD), Italia\\
           }

   \date{Received September 15, 1996; accepted March 16, 1997}

 
  \abstract
   {}
   {This work aims to advocate the benefit of high resolution spectroscopic monitoring in the study of local group transients. Taking advantage of the experience gained through a vast and systematic study of transients such novae and non-compact-stellar-mergers which we monitored through panchromatic high resolution spectroscopy, we show that a similar approach -i.e. a change of paradigm in the observing strategy and the analysis- in the study of a number of (local group) transients would be advantageous.  
   }
   {As an exemplary analysis, we focus on the optical transient NGC 300 OT2008-1.  Searching the ESO archives, we found a low resolution (FORS) and a high resolution (UVES) spectrum that were separated by only one day with no changes between them.
   We reduced and analyzed each spectrum comparing the different information accessible in the two cases. }
   {The independent analysis of the FORS and UVES spectra show that in the low resolution spectra we can securely identify only a small sample of lines and miss the correct characterization of the ejecta energetics which remain at the level of speculation. Instead, in the high resolution data, we  identify a larger sample of emission lines and analyze their profiles. This points to a complex geometry and ejecta dynamics whose inference are simply impossible in low resolution spectra. Line profile studies are not possible with low resolution spectra, and may lead to potentially misleading measures. Additionally, because of the limited information available from low spectral resolution data, the interpretation is compromised (e.g. FWHM measurement of unresolved lines), and prevented formulating any realistic physical scenario  other than  conducting parameter fitting to oversimplified, biased, standard  models. In this occasion only one epoch was available, but monitoring is fundamental to characterize the transient evolution. In particular, since, based on the UVES spectrum, we propose a new very specific scenario, we discuss how it could have been confirmed or dismissed thanks to a high resolution spectroscopic monitoring.  Low and high resolution spectra serve different but complementary purposes. With LR one does bolometric-like studies (SED, strengthening or weakening of the major transitions), while with  HR one does dynamics and precise physical characterization.  We also show examples from our past works, on different types of transients, for which it was possible, by monitoring with high resolution spectroscopy, to disentangle the various components, constrain their physical parameters, the involved energy source, and derive the ejecta dynamics. }
   {}


   \maketitle
%

\section{Introduction}
Spectroscopic observations of transients, in particular extra-galactic transients, are mostly limited to low resolution (LR) for a number of obvious reasons. Although these transients can be intrinsically luminous, they are distant so apparent magnitudes only occasionally reach 16th mag. Furthermore, stellar transients usually fade rapidly. 
Thus, broad-band photometry is generally the favored observing strategy since it permits long followup and is cheap, time-wise. In addition, theoretical models often predict only photometric behaviors. Optical spectroscopy with a resolution (R) of a few hundred is somewhat more demanding in terms of telescope time than photometry but still widely used for classification and long term monitoring; while, high-resolution (HR) spectra (with R up to several thousands) is considered prohibitive even with the current 10m-class telescopes. 

The initial exploration of the diversity of transients (when all objects were generically tagged as novae) was necessarily framed phenomenologically as a collection of events that have gradually been taxonomically distinguished on the basis of new observational criteria, including the peak brightness, the luminosity decline rate, the color evolution and, eventually, their spectral appearance. The emission line width, which is indicative of the expansion velocity of the gas, is one of the most immediate physical parameters to estimate. In this context, the extremely large velocities often exhibited by the spectral lines (thousands to tens of thousands km/s) promoted the routine use of LR spectroscopy since the line widths are already resolved with R of a few hundred. 

Furthermore, the requirement of target of opportunity (ToO) and/or multi-epoch observations in time-critical fashion, has often limited the monitoring to small or mid-size telescopes, as the pressure of the scientific community on these facilities is usually lower. As a consequence of these strategies, extended databases of light curves and LR spectra have been accumulated (e.g. ePESSTO and NUTS collaborations), and recently complemented by deep and high angular resolution archival images from space telescopes showing the transient's environment within its host galaxy. All of this allowed a statistical inference of their origin, including the basic properties of their progenitors and a tentative physical scenario of the outburst. However, in most cases, the ultimate interpretations are not unique and the actual physics remains somewhat obscure. This is even more evident when the transient event occurs while the star is still embedded in a high density, clumpy and asymmetric circumstellar medium (CSM). This is frequently observed in ejecta-CSM interacting supernovae (Smith 2017). A further complication is that a range of progenitor configurations and eruption mechanisms may produce stellar transients exhibiting surprisingly similar observational properties when the ejected material interacts with the CSM (e.g., Pastorello \& Fraser 2019). In all these cases, simplified scenarios inferred through only photometric and LR spectroscopic observations can be highly ambiguous or even wrong. 

Instead, the most significant progress always occurs when individual objects are sufficiently bright and/or close to be observed with high cadence, providing high signal-to-noise data and, most importantly, using complementary approaches to those usually adopted for normal transients, including (spectro-)polarimetric observations (e.g. Leonard et al. 2000; Desidera et al. 2004; Singh et al. 2024; Shrestha et al. 2025) and HR spectroscopy (e.g., Kaminski et al. 2009; Smith \& Andrews 2020; Smith et al. 2023; Pessi et al. 2024). 

In this work, we emphasize the importance of systematic HR observations of transients to determine the dynamics of the expanding gas and constrain the physics of the  ejection process and the transient's formation channel. This is crucial when the observed velocity is low (<1000 km/s) and/or there is a conspicuous amount of circumstellar gas. Such HR observations are limited to the Local Group of galaxies because of the currently available telescope collecting power.
An appealing test case is that of a peculiar interacting transient NGC 300OT 2008-1\footnote{This transient has recently been named with an IAU designation as AT 2008jd (Monard 2022).} (300OT from hereafter). This transient has been extensively studied in the past 15 years (e.g. Prieto et al. 2009; Bond et al. 2009; Berger et al. 2009; Gogarten et al. 2009; Thompson et al. 2009; Kashi et al. 2010; Patat et al. 2010; Ohsawa et al. 2010; Kochanek 2011; Humphreys et al. 2011; Soker \& Kashi 2012; Adams et al. 2016; Valerin et al. 2025a, 2025b). It is the prototype of a novel class of ejecta-CSM interacting gap transients labeled Intermediate-Luminosity Red Transients (ILRTs)\footnote{Data for other ILRTs were published by Prieto et al. 2008; Smith et al. 2009; Botticella et al. 2009; Wesson et al. 2010; Szczygiel et al. 2012; Cai et al. 2018; Cai et al. 2021; Jencson et al. 2019; Stritzinger et al. 2020; Karambelkar et al. 2023; Moran et al. 2024; Valerin et al. 2025a,b.}. While their light curves resemble type II supernovae, they are 1-2 orders of magnitudes fainter. Their spectra show narrow emission lines of H, and a characteristic [Ca II] doublet visible throughout the ILRT evolution. The progenitors have been observed in quiescence in mid-infrared imaging taken by the {\it Spitzer} spacecraft, and were moderate-mass stars embedded in dusty environments. The observables of ILRTs appear to be consistent with the expectations of electron-capture supernovae from super-AGB stars. The evidence that the progenitors of an handful of ILRTs have also been found to disappear several years after the outburst  in the mid-infrared supports their identification as terminal core-collapse explosion of the progenitors (Adams et al. 2016, Valerin et al. 2025a,b).

Our purpose in this paper is to demonstrate the impact of HR spectroscopy in the study of stellar transients. Our intention is not specifically to attempt a definitive scenario for a particular transient, but to present an example of how the chosen observing strategy constrains the interpretation. We thus analyzed a HR UVES a LR FORS spectrum of 300OT, independently of any already published result and without taking in account the transient classification. These two spectra were taken at the VLT only one day apart, a time interval sufficiently small to exclude any significant spectral development in an ILRT. They were retrieved through the ESO archive.\footnote{\url{https://archive.eso.org/eso/eso_archive_main.html}} (program: 281.D-5016, PI: Benetti). These data are presented in Sect. 2. We  analyze in detail the two spectra, comparing the amount of information contained in each of them in Sect. 3. In Sect. 4, we  emphasize the importance of 
multi-epoch observations spanning the evolution of the transient. In Sect. 5 we  discuss the limitations induced  by the reduced wavelength coverage. Finally, a discussion and the conclusions are presented in Sect. 6. 

\section{The VLT archive and the data reduction}
The UVES spectra were reduced with UVES pipeline version 6.1.3 and the graphic interface {\it gasgano}, version 2.4.8. The reduction process was sped up downloading the master calibration frames (an option which works with {\it esorex} but not with {\it esoreflex}). The flux calibration was performed using the master response function produced by the observatory\footnote{\url{https://www.eso.org/observing/dfo/quality/UVES/pipeline/response.html}}. The FORS spectra were reduced using IRAF's recipes\footnote{IRAF is distributed by the National Optical Astronomy Observatory, which is operated by the Association of Universities for Research in Astronomy (AURA) under a cooperative agreement with the National Science Foundation.}. The procedure was quite standard although the FORS data were taken with the polarimetric optics (a Wallaston prims and a retarder-wave plate - see the FORS User Manual\footnote{\url{https://www.eso.org/sci/facilities/paranal/instruments/fors/doc.html}} for more details) and the object light in each exposure was split in two beams of orthogonal polarization, the ordinary and extraordinary spectra.  We separately extracted and wavelength calibrated  these two spectra in each frame and then combined the two, flux calibrating their sum.  Finally we averaged the four spectra obtained at the four retarder wave plate angles (0, 22, 45 and 67 deg), also verifying that they were comparable within the uncertainties. For the flux calibration we used the spectophotometric standard star CD-32~9927 which was observed $\sim$1 month in advance with the same instrument setup (GRIS300V and no order sorting filter). FORS GRISM300V was used without order sorting filter and, therefore, is affected by order overlap from about 6600 \AA \ red-ward. The same wavelength range is also strongly affected by fringes which we tried to remove using the spectrophotometric standard\footnote{After a number of tests, we chose not to use the flat field in the whole reduction process since, because of its odd illumination, it introduces artifacts at the continuum level which is very hard to remove.}. The apparent emission features beyond $\sim$8700 \AA \ are likely residual fringes (the fringe pattern in the science spectrum and the standard spectrum are not exactly overlapping because of the different instrumental flexures) and are, therefore, ignored. 
We note that there is a slight discrepancy between the two final spectra in terms of flux calibration. The color dependent part is probably ascribed to the fact that FORS uses the telescope LADC (longitudinal atmospheric dispersion corrector, Avila et al. 1997) which is set at the beginning of the acquisition and does not change automatically (with airmass) as the observation proceeds (section 3.11.4 of the FORS User Manual, issue 116). Instead, the UVES slit rotates with the sky and, by default, is set along the direction of maximum dispersion caused by the atmosphere. The spectra could be corrected with existing photometry, if needed. 

\section{Low versus high resolution spectroscopy}\label{lrhr}

   \begin{figure}
   \centering
   \includegraphics[width=9cm]{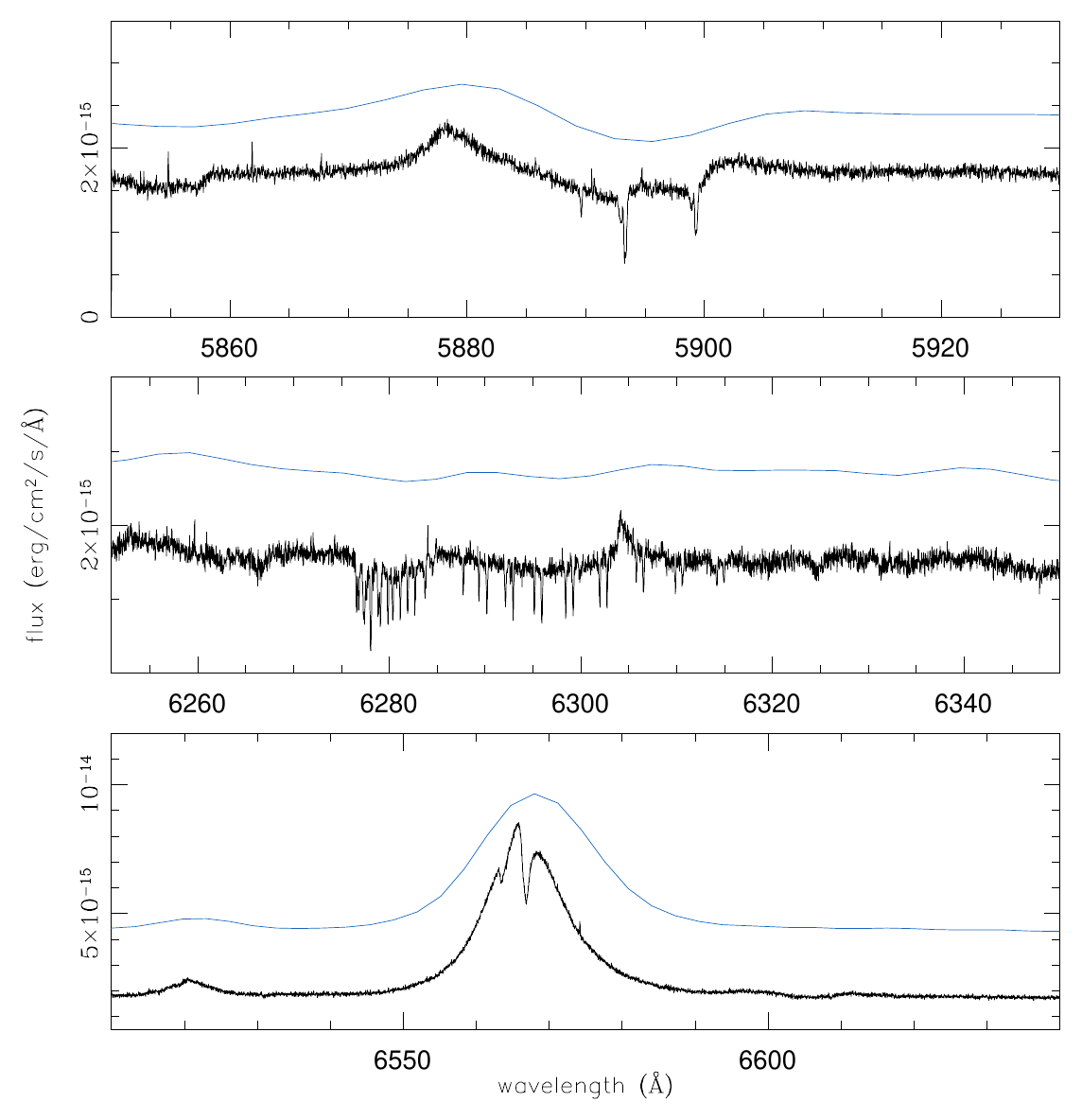}
      \caption{"Smear-out" effect and inaccurate wavelength solution of the LR spectroscopy (blue line, FORS spectrum) compared to the HR one (black line, UVES spectrum). See Section \ref{lrhr} for more details. In the bottom panel, the FORS spectrum has been offset by the constant +2E-15 erg/cm$^2$/s/\AA \ for clarity. 
              }
         \label{fig2}
   \end{figure}
The difference between HR and LR may seems obvious and well known but with the widespread use of LR spectroscopy among the transients community, the limitation intrinsic to LR seems quite neglected. Consequently, it is worth stressing two singularly important aspects of the spectroscopic analysis: 

$\ast$ {\it The process of line identification is linked to the completeness of the sample of available lines}: the larger the number of detected (absorption and emission) lines, the more trustworthy the identification is, with all that this reliability implies. A single line cannot be securely identified and produces only best guess. A line identification is solid when multiple transitions from the the same multiplet and/or ion are also detected. LR spectroscopy, by smearing out (i.e. loosing) weak transitions and blending transitions of close wavelength, jeopardizes the line identification process, limiting it to only a few species. In addition, unresolved blends, together with the larger uncertainty of the wavelength solution, might lead to the wrong identification. Fig.~\ref{fig2} demonstrates how the weak narrow features, in emission and absorption, are simply lost. From top to bottom, the Na~I interstellar absorptions, the forbidden [O~I]$\lambda$6300 emission together with the water atmospheric absorptions, and the narrow absorption superposed on the H$\alpha$ emission are not present in the LR spectrum. 

\begin{figure*}
   \centering
   \includegraphics[width=9cm]{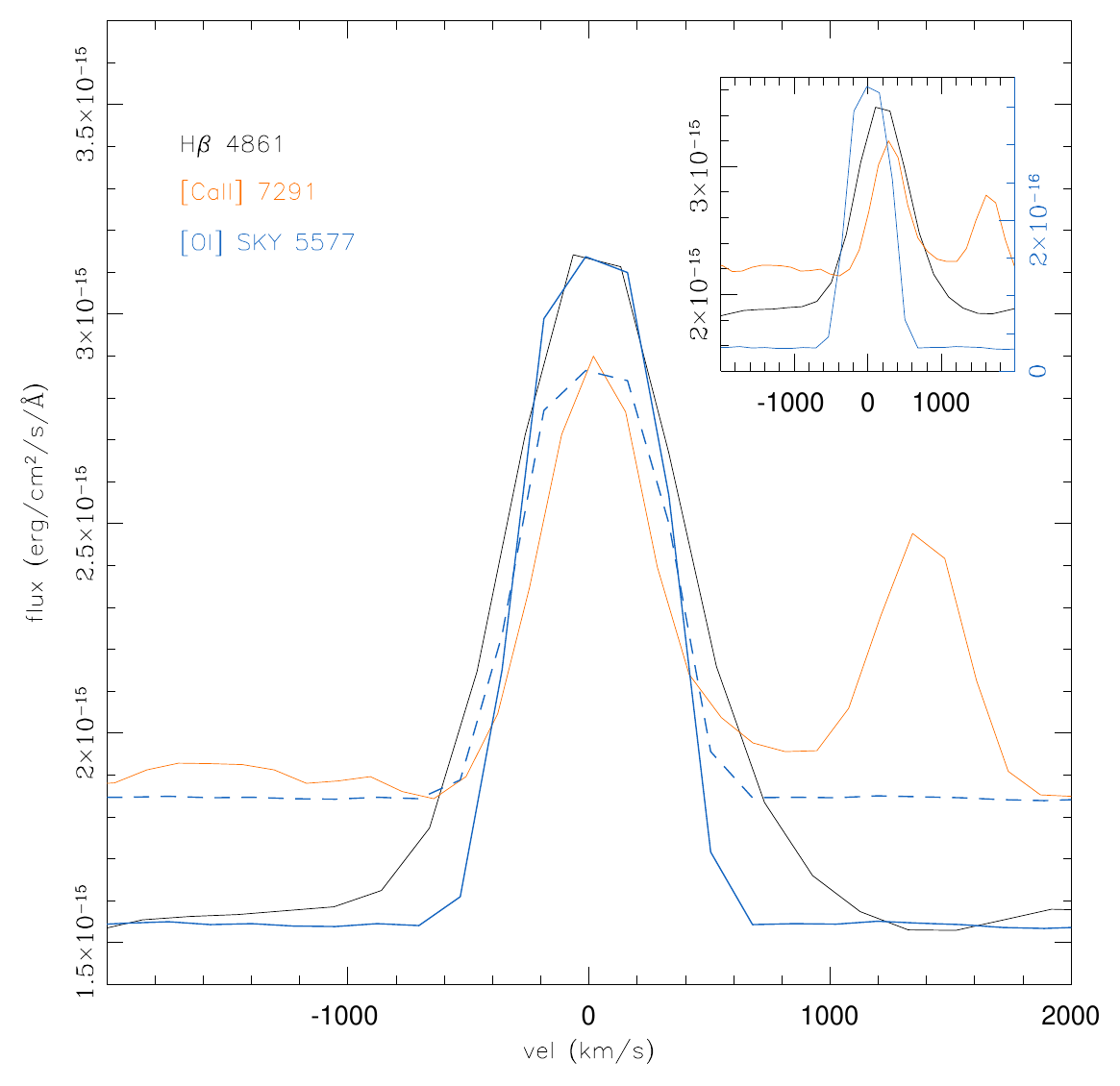}
   \includegraphics[width=8.55cm]{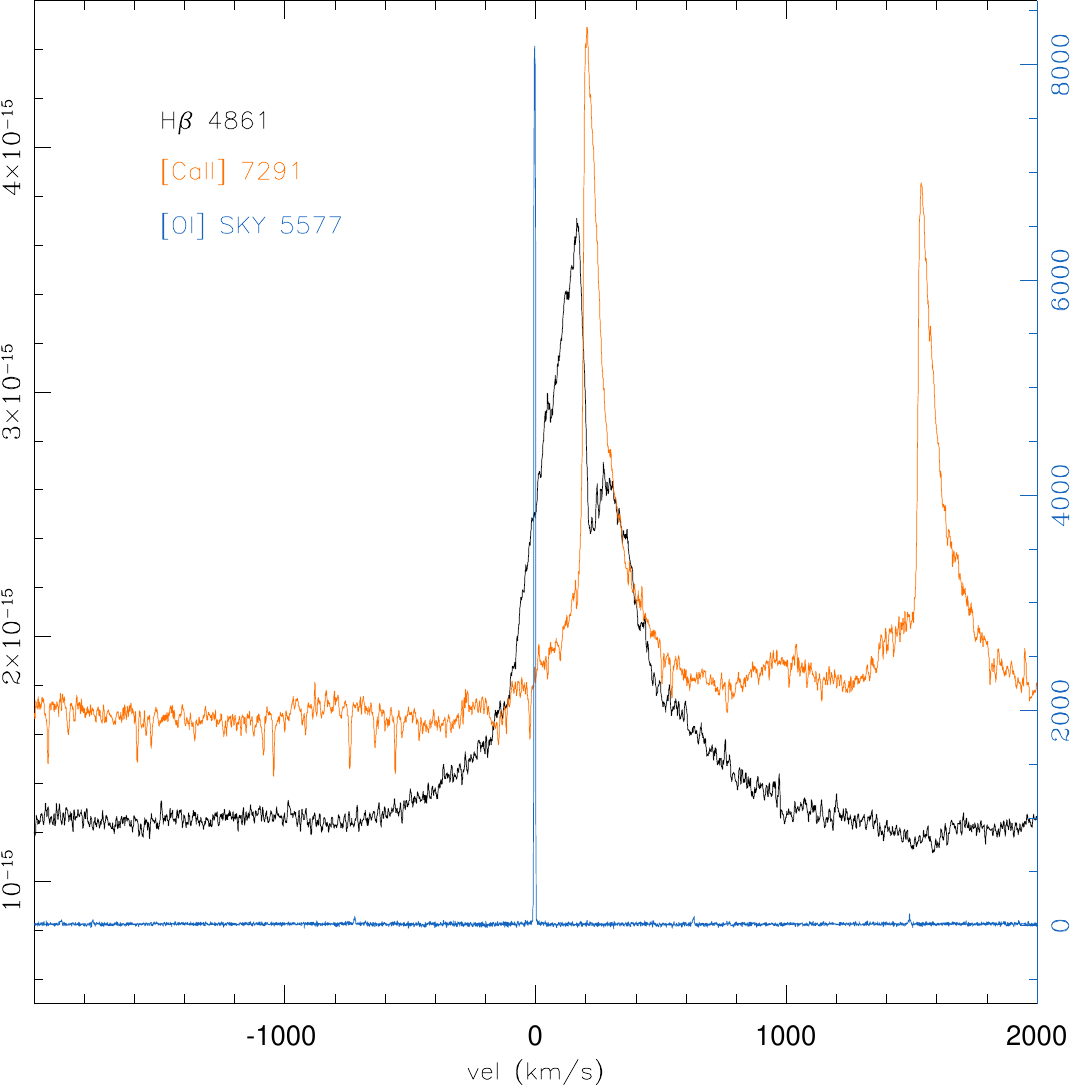}
  \caption{Line profiles comparison between the transient's H$\beta$ and [Ca~II]$\lambda$7291 and the sky emission [O~I]$\lambda$5577, in the FORS (left panel) and UVES (right panel) spectra. In the left panel, the [O~I] sky emission has been scaled to the H$\beta$ (blue solid line) and the [Ca~II] emission lines (blue dashed line), to better show that the H$\beta$ wings are resolved, while those of the [Ca~II] emission are not. In addition, the H$\beta$ and [Ca~II] emission lines have been offset (by -3.0E-16 and -3.2E-16 erg/s/\AA/cm$^2$, respectively) and shifted in velocity (by -168 and -260 km/s, again respectively) to facilitate the comparison (NGC 300 heliocentric redshift is 147 km/s, Kachaeov et al. 2018).  Their original alignment and relative flux is shown in the inset.  
  In the right panel (UVES spectrum), both the H and the [Ca~II] lines are clearly resolved. The [OI]$\lambda$5577 sky emission is in ADU (right y axis).
  }
         \label{fig4}
   \end{figure*}

$\ast$ {\it The line profile analysis is linked to dynamics, separating it from the energetics}. Line profile analysis is, however, not possible in LR spectroscopy. By comparing profile structures, it is possible to strengthen the identifications, and to distinguish different line forming regions. This provides information about the geometry and the stratification of the emitting gas. In LR spectra the lines, typically fitted with Gaussian and/or Lorentzian profiles, are dominated by the spectrograph's line spread function, LSF, leading to false dynamical information. Fig.~\ref{fig4} compares the profiles of the H$\beta$ and [Ca~II]$\lambda$7291 emissions from the transient with the [O~I]$\lambda$5577 from the sky (a tracer of the spectrograph's LSF) in the FORS and the UVES spectra. In particular, in the FORS spectrum (left panel) the [Ca~II] line is unresolved, while the H$\beta$ emission is only marginally resolved. Instead, in the UVES spectrum, both the H$\beta$ and [Ca~II] emissions are resolved and are fundamentally different in morphology, width and spanned velocity range.

In the following subsection we analyze the LR and the HR spectra independently, to further underline the different reach of the two strategies.

\subsection{The two spectral analyses at face value: results from the FORS spectrum alone}

The FORS spectrum, with its superb S/N$\sim$200-400, allows robust identification of the H Balmer lines down to H$_8$, H-Paschen 10 and 11 (the latter in blend), He~I 5875, the [Ca~II] doublet and the Ca~II NIR triplet, in emission, Na~I~D and CaII~H\&K in absorption. Other weaker features are visible but their identification remains uncertain. As typical for most LR grisms, the wavelength solution is not very accurate and is also affected by a nonlinear component most evident at the blue and red ends of the spectrum. To this is added the smearing effect of the low resolution. Thus, lines from the same series or multiplet have different wavelength accuracy, and even after redshift correction they display a range of offsets with respect to rest. For example, the hydrogen Balmer lines have measured offsets (after redshift correction) in the range -2.0 to +0.7 \AA.  Additionally, the red He~I lines have offsets of -0.7 to 1.2 \AA.  The lines measured at 5018.9 and 4925.7 are more consistent with Fe~II RMT42 $\lambda$5018.4 and $\lambda$4923.9, than He~I $\lambda$5015.7 and $\lambda$4921.9, but the Fe~II RMT42 $\lambda$5169.0 seems missing. Of the attempts to identify the blend peaking at $\sim$5174 \AA, none is convincingly close to the Fe~II $\lambda$5169.0 so that the presence of Fe~II RMT 42 cannot be confirmed. For similar reasons, a number of weak emission features remain unidentified or highly uncertain (e.g. He~II $\lambda$4686) and, with that, the ionization state of the emitting gas is uncertain. 
In short, line identifications are possibilities rather than certainties with LR spectra and the transient's energetics are only constrained by its luminosity. Lacking P Cyg profiles, the expansion velocity would be estimated by the width of the emission lines. The only other physical information that can be obtained about the line forming region is that the density of the emitting gas is low, given the presence of the [Ca~II] doublet. Had this spectrum been taken at the typically much lower S/N, e.g. $\sim$20-30, the number of identified lines would have decreased by 30-40\%. 

In both the low and the high S/N spectra, the detected transitions are inevitably ascribed to a same region, whose nature could be virtually anything (precursor, ejecta, etc). In this case, the combination of the observed low density (from the [Ca~II] lines), the steep Balmer decrement H$\alpha$/H$\beta$ ($\sim$4) suggesting a high optical depth and high column density, and the small/unresolved line width, would have favored the precursor interpretation. 
A sequence of spectra of similar quality taken across the outburst would have allowed measurement of only the relative weakening/strengthening of the various transitions. 

\subsection{The two spectral analyses at face value: results from the UVES spectrum alone}\label{uves_a}
Compared to FORS, the UVES spectrum has similar wavelength coverage and substantially lower S/N ($\sim$5-30). Nevertheless, what appears as a wavy continuum in FORS is actually a number of weak emission and absorption features. These features crowd together in the blue part of the spectrum, making the line identification difficult, but clearly showing that there is a "curtain" of absorbing metals intercepting the U (and the NUV) flux. In addition to the dozen lines identified in the FORS spectrum, we confirm the presence of Fe~II (of which particularly strong are multiplet 42, 40, and 74, 48 and likely 49), Ti II (multiplets 41 and 31). Other Ti II multiplets are blended with other metal absorptions and not easily distinguishable. The same applies to Sc II (solid evidence of multiplet 29 only). Also confirmed is the presence of OI (multiplet 1, 4 and IF). He~I is detected, in particular $\lambda$5875.6, 6678.1 and 7065.2.  The presence of He II can be dismissed despite lacking wavelength coverage around $\sim$4686 \AA \ and 3203 \AA, since we do not detect the He II$\lambda$10123 line, or the weaker but relatively isolated $\lambda$5411. Transitions from other high ionization potential energy elements are also undetected. Therefore, we conclude that the gas is relatively cool and dominated by transitions from singly ionized metals. Most importantly, we can also state that there is UV-NUV radiation heavily absorbed by the optically thick resonant transitions of the same metals (iron peak elements) which de-excite  through alternative channels producing the emissions in the optical range. Hydrogen and Helium are (singly) ionized and recombine, while the oxygen is mainly excited by fluorescence (see below). 
Despite its relatively low S/N, the HR UVES spectrum allows for a better characterization of the ionization state of the emitting gas than the LR FORS spectrum, regardless of its S/N. 

However, the major advantage in the use of HR is in the line profile analysis which is impossible in the LR mode and which produces information that constrains, if not pinpoints, the underlying physical processes. 
   \begin{figure}
   \centering
   \includegraphics[width=9cm,angle=0]{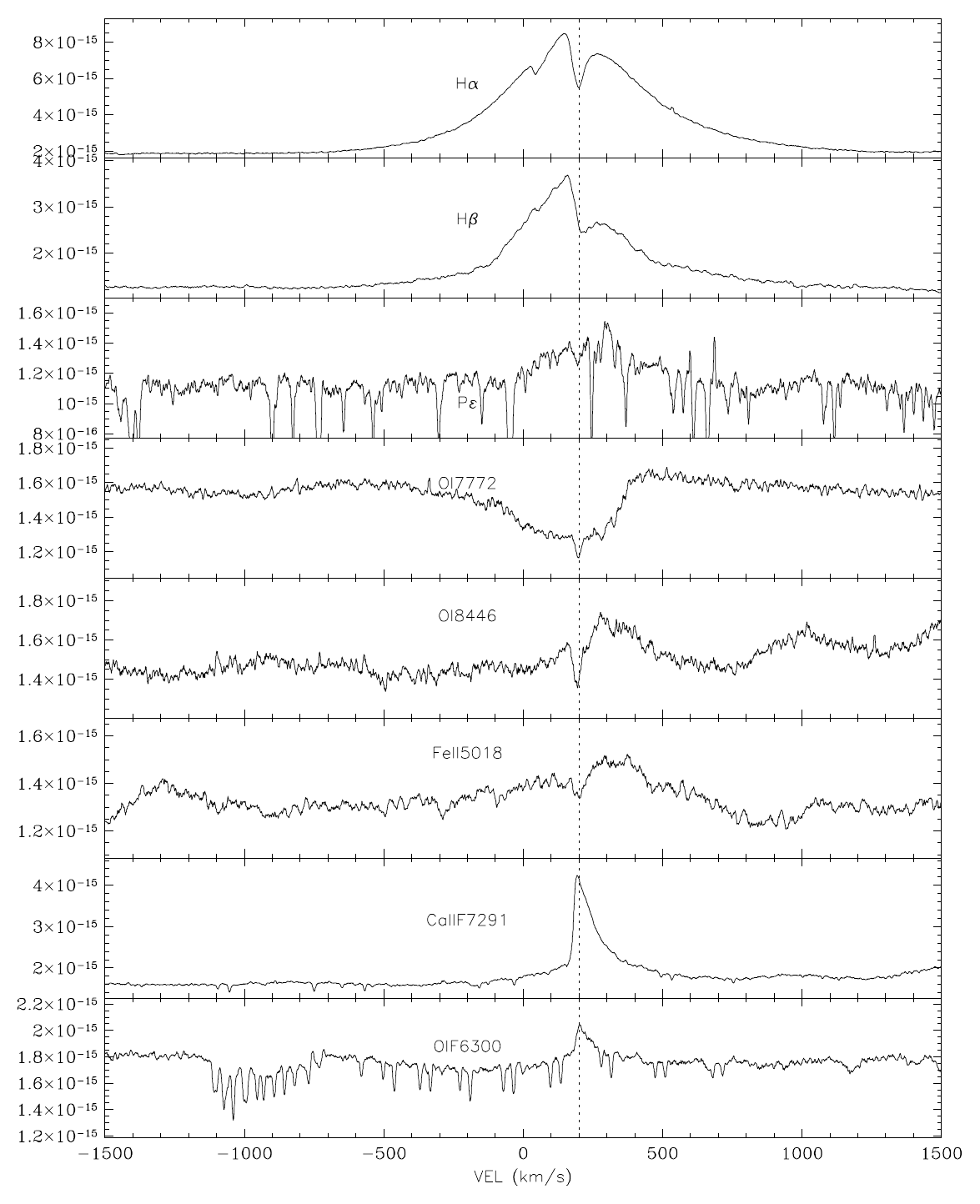}
  \caption{Selection of transitions showing the narrow absorption component mirrored by the [Ca~II] and [O~I] emissions (see text for more details). The spectrum has been smoothed with a boxcar smoothing function of width 15 points. Y-axis unit is Flux in erg/cm$^2$/s/\AA. The vertical dotted line marks the +200 km/s velocity roughly matching the minimum of all the observed absorptions. 
              }
         \label{fig5}
   \end{figure}
The most striking feature in the HR spectrum that is invisible at LR is a narrow absorption at about +190 to +200$\pm$10 km/s (LSR reference frame, spectrum not corrected for the host galaxy redshift). As shown in Fig.\ref{fig5}, this absorption is common to several transitions (H Balmer and Paschen, a number of Fe II transitions, Ca NIR, and O I). It is asymmetric, being very sharp/steep in its blue side and with an extended wing on the red side, and is replicated, in profile, by the emissions in the forbidden transitions from O~I and Ca~II and perhaps Fe~II (only multiplet 14F could be ascertained). Because of the nature of the transitions (from H and excited level of ionized metals) in which it is observed, it clearly originates in circumstellar gas. Only the Na~I~D absorption at a similar velocity cannot be firmly attributed either to the host galaxy or the transient's circumstellar environment. Most likely, given that the Na~I~D narrow absorption is composite, it arises from both. The circumstellar gas responsible of the narrow absorption component is only weakly ionized (metals) and has a relatively low density, not much larger than several 10$^6$ cm$^{-1}$ (e.g. Dessart and Hillier 2020, otherwise the O~I and Ca~II forbidden transitions would not appear). In addition, since the absorption is detected also in the Paschen series (although weakly), both Ly$\alpha$ and Ly$\beta$ must be very opaque to substantially populate the n=2 and n=3 levels of the the hydrogen, respectively, and the gas column density -in the narrow component- must be large. The high opacity of Ly$\beta$ also explains the fluorescence of the O~I triplet $\sim\lambda$8446 and the detection of its narrow absorption components.

This absorbing component is completely missed in the LR spectrum. It is only detected, in emission, in the forbidden Ca~II multiplet, but it is kinematically undistinguished from all the other emission lines (which we will call "broad" in the following) and, therefore, everything is ascribed to a same uniform  circumstellar matter (CSM) in view of the measured width ($<$1000 km/s). But the gas producing the narrow absorption and the forbidden Ca~II and O~I emissions is a separate region from that producing the broad emission against which it is detected. The kinematics of these regions are different and so is their degree of ionization/excitation, the broad component showing He~I which is not observed in the narrow component. 

The kinematics of the narrow component is particularly informative. NGC 300 is a spiral galaxy with an intermediate inclination and an average redshift of 136 km/s (with respect to the LSR, Westmeier et al. 2010), but its outskirts or spiral arms will have some additional radial component. From the 2D mapping of Westmeier et al. 2010, the radial velocity at the position of 300OT is in the range 181$\leq$v$_r\leq$196 km/s (with respect to the LSR) with a dispersion of only 12 km/s. 
Now, the narrow (absorption) component has its minimum around 180 to 210 km/s suggesting a gas region moving away from the observer and, possibly, toward the transient. At the same time, it has a peculiar asymmetric profile indicative of a gas kinematic that is not typically observed from gas being accreted by the central object (e.g. Grinin et al. 1994, Folha \& Emerson 2001, Batalha et al. 2001, Perez et al. 2008, Hamilton et al. 2012, Cauley et al. 2015, Hajigholi et al. 2016, Akimoto \& Itoh 2019, Yang et al. 2022, Erkal et al. 2022). The narrow absorption component can be explained as the neutral gas (not yet affected by the transient ejecta) from a donor companion wind, similarly to the circumbinary material produced by the giant's wind in symbiotic recurrent novae (SyRN, e.g. Shore et al. 2011a, Walder et al. 2008, Orlando et al. 2009). The orbital motion accounting for its redshift (or null velocity) with respect to the transient' systemic velocity and orbital phase at the time of the observations being in the first elongation or the first two quadrants (i.e. the donor and its wind are moving away from the observer). Then, similarly to the dynamics observed in SyRN in nova outburst, the broad component matches the shock front between the ejecta and the circumtransient/circumbinary material (Shore et al. 2011a). 

Considering the Roche lobe geometry for a generic binary where the outburst occurs in one of the stars, the expanding ejecta will smash directly against the companion wind in the direction toward L1 and the companion, but will travel with the wind (just at a higher velocity) in the opposite direction toward L3. The companion wind and the ejecta will impact at an angle in all the other directions. Similar impacts "at an angle" will occur above and below the binary orbital plane. Hence, the shock region is clearly asymmetric around the transient (no matter whether the ejecta have spherical symmetry) and with a range of velocities that together will produce a smoothed bell-shaped line profile. The steepness and the extension of the emission line wings will result, in addition to the energetics of the shock, from the observer's line of sight. The presence of a shock precursor will depend on both the (dynamically undisturbed) gas opacity and the observer's line of sight. 
The distribution of the circumbinary gas might be axisymmetric with a higher density in the orbital plane due to the conservation of the angular momentum. Since there will be different gas densities on the front and rear side of the outbursting object as well as close to or away from the orbital plane, it is reasonable to expect a range of gas conditions sampled across the line profile. 

   \begin{figure}
   \centering
   \includegraphics[width=9cm,angle=0]{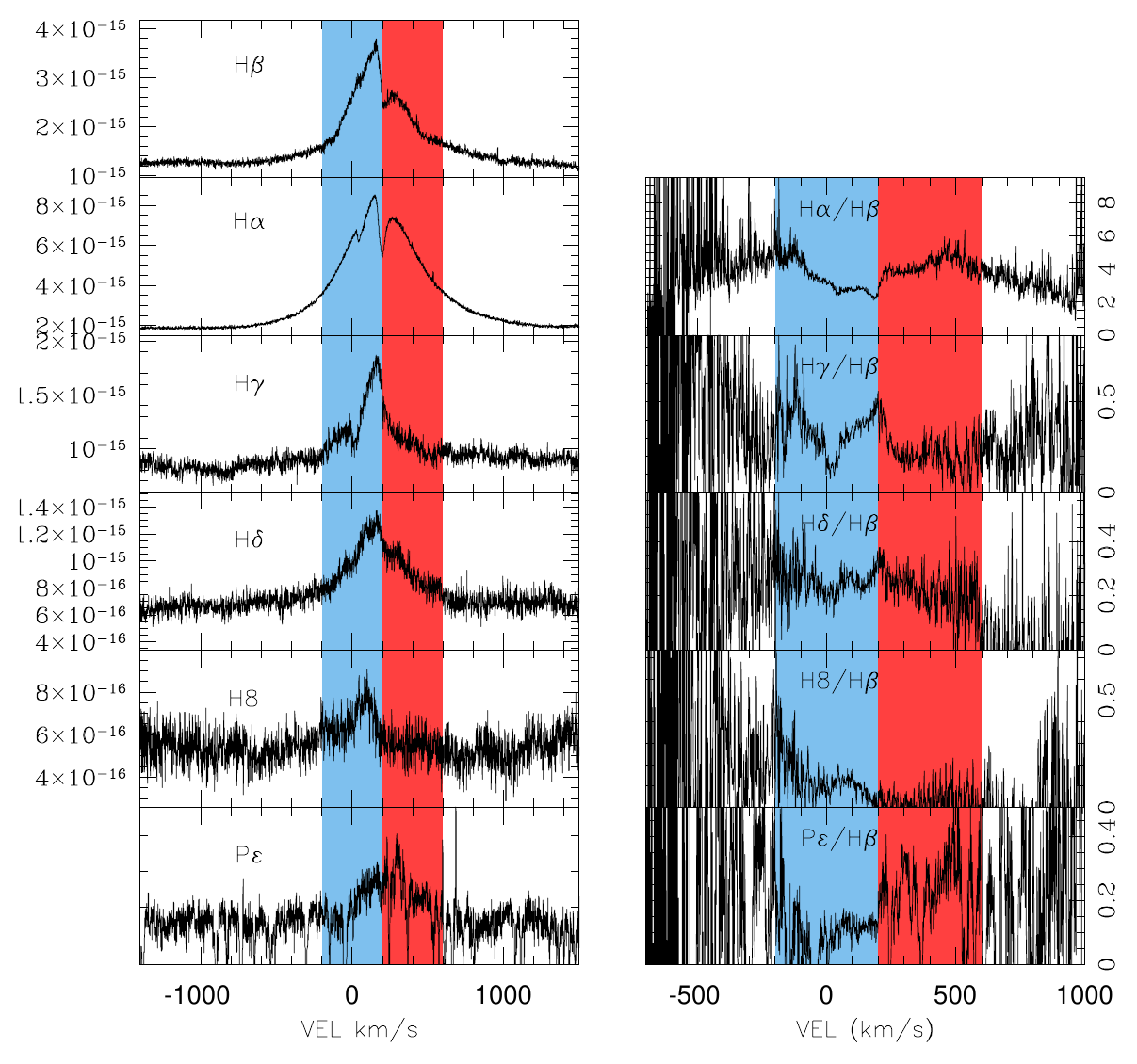}
   \caption{Left Panel: H Balmer and Paschen~$\varepsilon$ line profiles. H$\varepsilon$ of the Balmer series has been omitted since is severely blended with the Ca~II$\lambda$3968 line (which is dominating). Right panel: the line ratio of the H lines over H$\beta$ in velocity space. Note the different velocity range adopted for the the left and the right panels, due to the noise in the line ratio at large velocities. The blue and red shaded area, together, delimit the core of the broad emission line and represent the part where the difference in the ratio is significant and robust. The wings, having less signal, do not permit equally solid inferences, although a difference is apparent. See text for more details. }
         \label{fig6}
   \end{figure}

We can check for those by computing the Balmer decrement (with respect to H$\beta$) per velocity bin (Fig.\ref{fig6} right panel; the left panel shows the individual line profile to aid the reader). Fig.\ref{fig6} shows a flatter decrement on the blue side of the emission component with respect to the red side, suggesting larger opacity for the latter. 
The figure also shows that the wings tend to behave differently from the line core, being in particular less opaque. Hence, either we are dealing with a medium consisting of multiple components (two to four, significantly complicating the geometry and the physics of the event) or with one characterized by a continuum of different/varying physical conditions as in the case of a shock front propagating in a CSM with a gradient (not homogeneous).  
None of this could be inferred from LR spectra. 

   \begin{figure}
   \centering
   \includegraphics[width=9cm,angle=0]{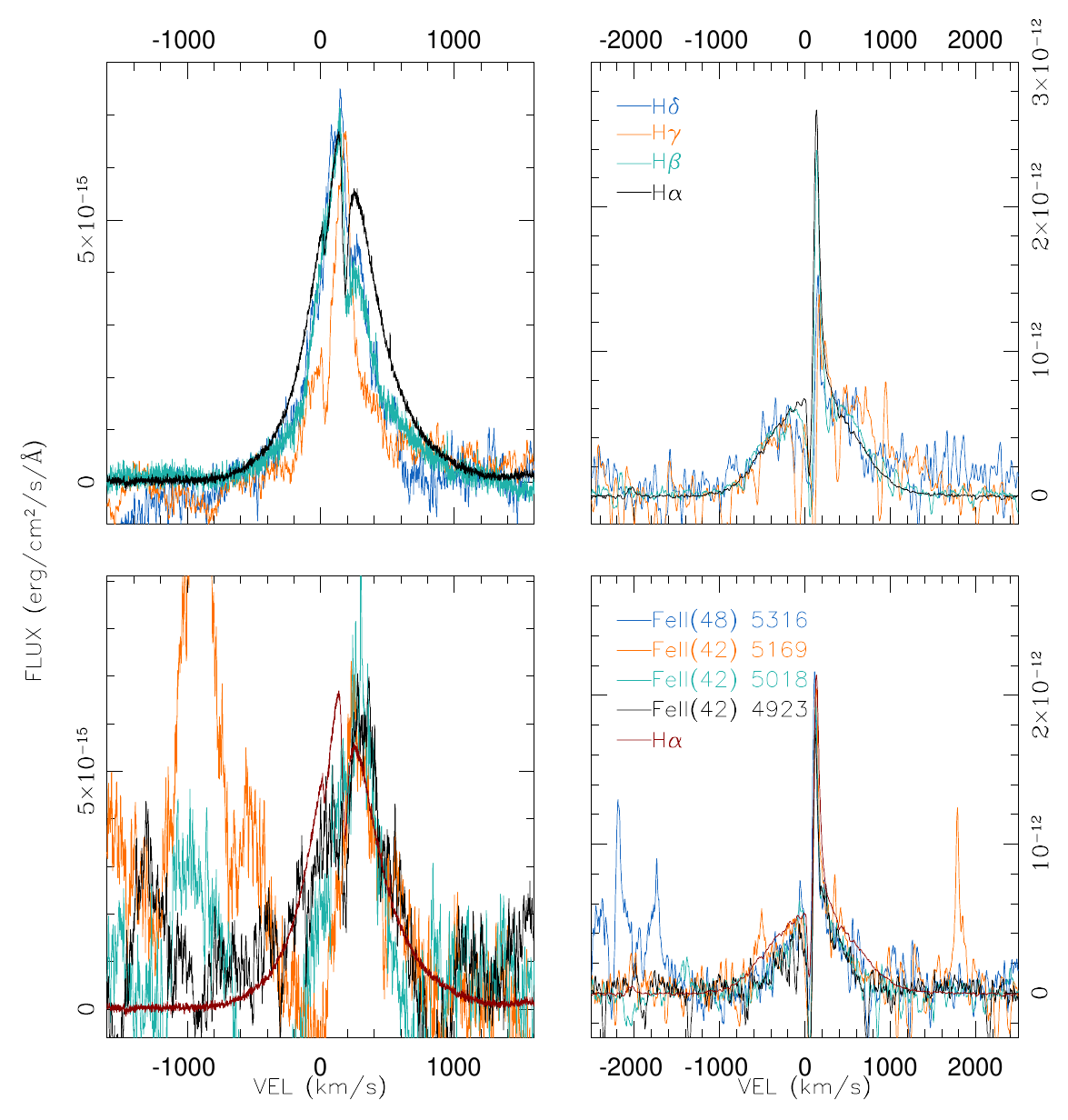}
  \caption{Scaled profiles (broad component) of the H Balmer (top) and the Fe~II (bottom) lines to the H$\alpha$ profile. Left panels: the 300OT case showing weak evidence of electron scattering. Right panels: the symbiotic nova V1413 Aql (ARAS database, observer: Francois Teyssier), whose wings are indicative of electron scattering.  See text for more details. 
  }
         \label{f9b}
   \end{figure}
The wings deserve further attention since they could indicate electron scattering, often observed (and expected) when shocks are present. Although (e.g. Fig.\ref{fig6}, left panel), the wings of the Balmer emissions show different extension, suggesting that dynamical effects are dominating over electron scattering, they might result from the low S/N in the higher line of the series. It is essential to collect high S/N observations of the Balmer series to establish the presence of electron scattering. The comparison with transitions from other species helps but might be ambiguous since different emission measures and line formation mechanisms have to be taken into account. 
Fig.\ref{f9b} shows in the two top panels the H Balmer lines H$\alpha$ to H$\gamma$ scaled to the peak intensity of the H$\alpha$  broad component for 300OT (left) and the symbiotic nova V1413 Aql (right).  In the bottom panels the same exercise is repeated with a selection of Fe~II emission from multiplets (RMT) 42 and 48. The matching among wings profiles in the case of V1413 Aql is evident, less so in the case of 300OT. Therefore there is no conclusive statement for the latter, with the caveat that the S/N, blends, and scaling biases all affect the inference.  

   \begin{figure}
   \centering
   \includegraphics[width=9cm,angle=0]{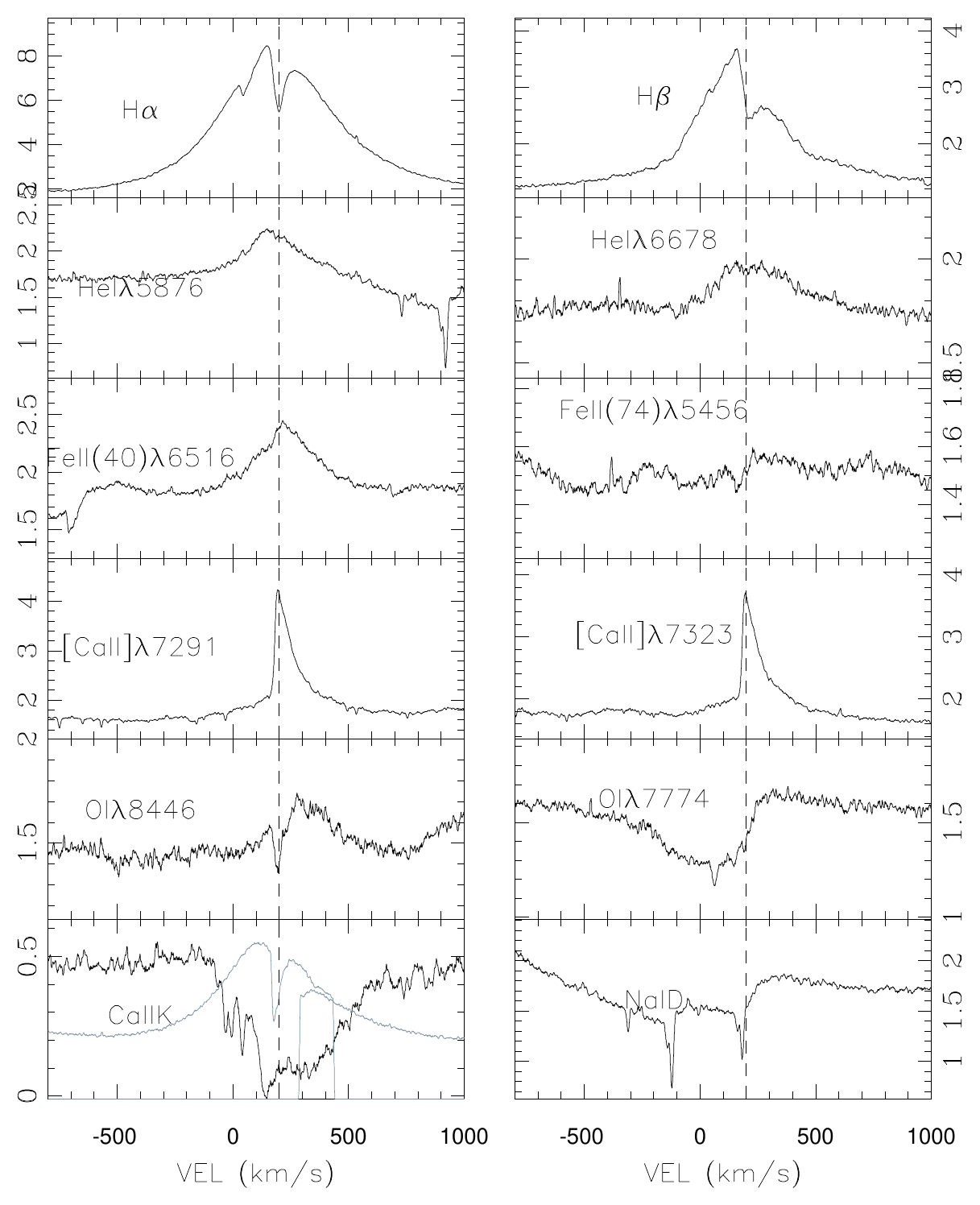}
  \caption{Comparison of the broad component profiles among various transitions showing how He~I and the Ca~II NIR triplet lines peaks on the blue side of the vertical dashed lines, while Fe~II, O~I, and Na~I peak at or on the red side of the same lines. See text for more details.  
  }
         \label{fig8}
   \end{figure}
Another interesting characteristic of the broad line profiles is that transitions from singly ionized low ionization potential elements (e.g. Fe~II, Ca~II), together with transitions from neutral elements (O and Na), peak at greater redshifts than the He~I transitions (Fig.\ref{fig8}). This suggests a gradient in the physical conditions of the medium producing the broad emission component. A velocity gradient tied to a stratified ionization  will produce different profiles and mean velocities for any non-spherical geometry. It is not the same for a spherical geometry since the radial velocity, $v_{rad}(r)$, the density, $\rho(r)$, and the ionization are only one dimensional. The inferred gradient and asymmetries are consistent with the proposed environment produced by the wind of a companion star where the shock front is propagating into. A single expanding medium would require a complicated velocity field and geometry to reproduce the same observations. 

The stratification of the circumbinary gas is apparent also from the analysis of the O~I and Ca~II transitions. The broad emission of the O~I triplet at $\sim$8446 \AA, together with the weakness of the O~I $\sim$7774 \AA \ emission, confirm the mechanism of fluorescence from the highly opaque Ly$\beta$. However, the broad absorption at the O~I triplet $\sim$7774 \AA \ indicates that other processes are occurring (absorption and collision), favoring the population of the $^5$S$^o$ lower level of the triplet. Most likely, regions that appear superposed in velocity space are distinct in the 3D geometry of the circumbinary gas and dominated each by a different radiation process. This also  explains the slight difference of the peak of the [Ca~II] broad emission and the Ca~II H\&K absorption with respect to the Ca~II NIR triplet (which is bluer).  

Both the narrow and the broad components indicate a high optical depth since they sample  different portions of the same circumstellar material: one (the broad) hit by the ejecta (i.e. dynamically interacting), and the other (the narrow) still dynamically undisturbed and simply absorbing the radiation coming from the shock region. The former is relatively small and  denser, the latter is much more extended and diluted into the circumstellar space. Although there is a density gradient, the overall column density is large. This explains the high  opacity of both components and the lack of an observed precursor emission. 

\section{The limits of a single (HR) epoch: the vital importance of a monitoring}
We have shown how a new physical plausible scenario that accounts for all the observational facts can be inferred from a single high resolution spectrum.   
Although this is only one possible interpretation, it does not require a complex and {\it ad hoc} geometry and/or formation history for the line forming region (both narrow and broad component). To confirm any  such scenario requires multi-epoch observations (i.e. time domain HR spectroscopy).  

\begin{figure}
   \centering
   \includegraphics[width=9cm,angle=0]{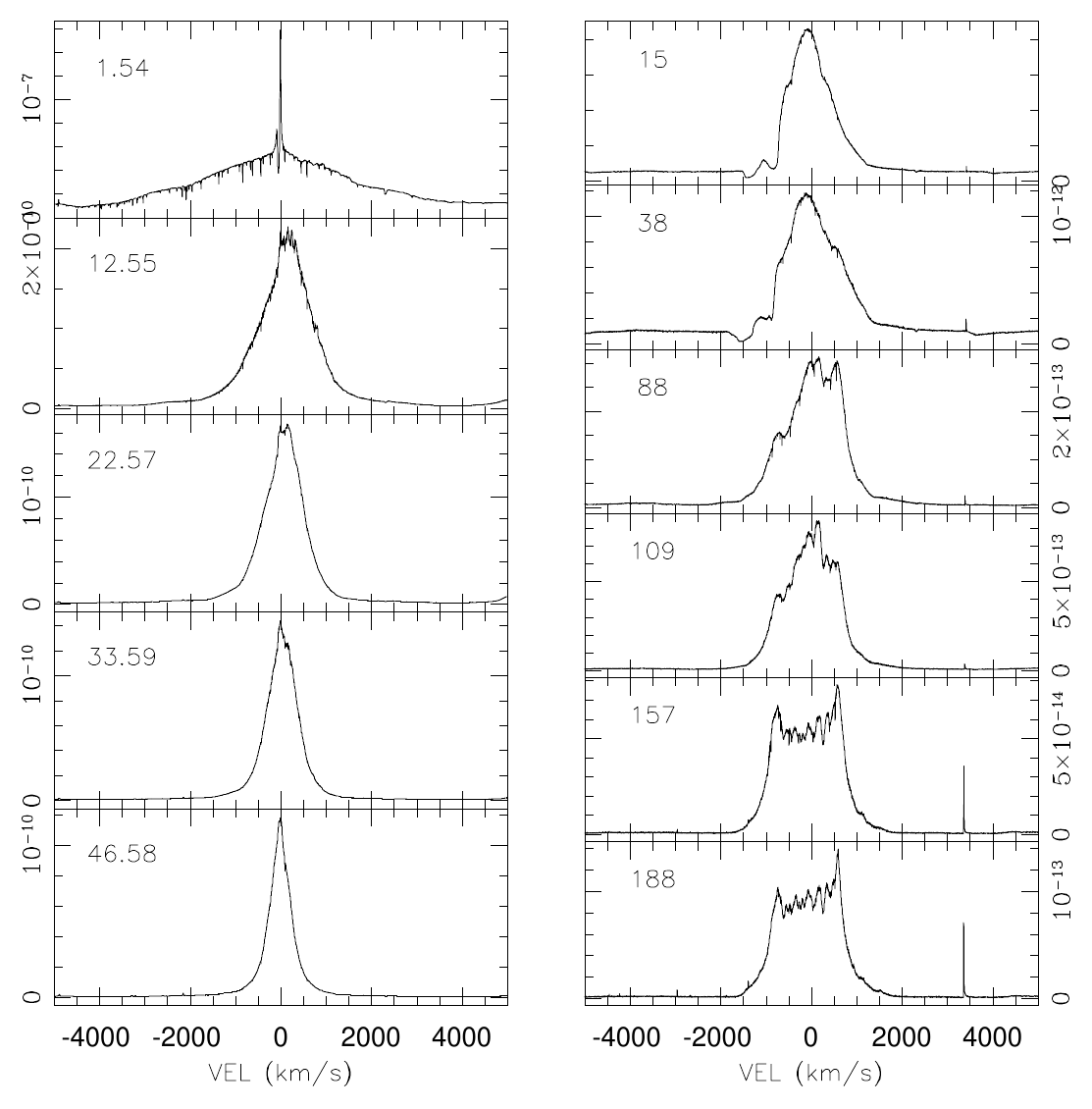}
  \caption{Examples of line profile evolution in the case of an ejecta expanding in a medium and producing shock emission in the symbiotic recurrent nova RS~Oph (left panel), and of ejecta in free expansion as in the classical nova V378 Ser (right panel). The two sequences were obtained at the 2.2+FEROS (R$\sim$48000).  
  }
         \label{fig9}
   \end{figure}

With time, the line profiles from a shocked gas change.  A precursor disappears, if initially present. The dynamically undisturbed neutral gas of the CSM (the narrow component we observe)  can also disappear, once the shock has crossed the whole medium and broken out. The profile of the broad component (the shock front) narrows, changing in shape or (a)symmetry depending on the shock breakout and the observer's line of sight. Different species might appear (e.g. coronal emissions) that display completely different line profiles since they form in the post-shock bubble. The sequence of H$\alpha$ profiles of RS Oph during its 2006 outburst (Fig.\ref{fig9} left panel) shows such a development: the precursor and the narrow absorption has disappeared by day $\sim$12 (with a velocity $\geq$4000 km/s the ejecta reach the giant companion in about half a day); the broad component has narrowed from FWZI$\sim$8000 km/s to $\sim$2000 in about 1.5 months from the outburst when the coronal emission such as [Fe~X] appeared. Coronal emissions have been reported for similar systems such as V407 Cyg and showed to be valid shocks diagnostic, in view of the strong correlation between the X-ray and [Fe~X] emission (Shore et al. 2011a, 2012a).

Conversely, optically thin ejecta in free expansion would maintain the line profiles while only decreasing in flux and narrowing in FWHM as the gas emission measure decreases due to expansion (e.g. the two bottom profiles in the nova Ser 2005/V387 Ser H$\alpha$ series in Fig.\ref{fig9} right panel; see also the gallery of nova line profiles in Mason et al. 2018, and Shore et al. 2016, 2013). During the initial optically thick phase, the line profile is constantly changing according to the gas geometry, homogeneity and the instantaneous location of the pseudophotosphere (first four profiles in Fig.\ref{fig9} right panel). Note how the absorption features that are visible in the first two epochs subsequently turn into emission. In addition, with just a single spectrum from, say, day 15 or 38, one would have inferred a P Cyg profile and a wind, an even more likely deduction from low spectral resolution spectra. 

Further, the persistence of any CSM absorbing component constrains its distance from the transient and therefore the CSM geometry and its previous history. Similarly, the increasing intensity (with no other line profile changes) of a photoionized CSM depends on its distance, the radiation diffusion time within the gas and, therefore, on the gas physical conditions, the duration of the ionizing source, and depending on ionization, the SED of the ionizing source (see e.g. Dwek and Felten 1992, for the continuum, and Mason \& Shore 2022, for line fluorescence).  Light echos can contribute additional emission components; the time of their appearance constrains the distance of the scattering echoing cloud.  All these processes can be distinguished through adequate monitoring\footnote{The definitive confirmation of a light echo, however, requires complementary linear spectropolarimetry.}.  
It is only the observed evolution that allows the distinction. 

In summary, changes in intensity and profile shape of different lines constrain the ongoing dynamics; while appearance or disappearance of different species informs about ionization changes and, therefore, about the powering source. If we were to have a sequence of HR spectra showing the profile evolution of the broad emission components and the persistence or disappearance of the narrow absorption component, or the appearance of new species or ions, we would have been able to produce more physical constraints in place of a merely plausible scenario. Below we describe a few cases for which the monitoring in HR spectroscopy lead to important physical constraints. 

HR spectra of the T~Pyx 2011 outburst early evolution showed that the apparent outward motion of the absorption component in the ejecta was consistent with an outward propagating  recombination wave (Shore et al. 2011b, in particular, their Figure 2, 4, and 5; see also Shore 2012b). In addition, those observations, combined with late time monitoring including the HR UV wavelength range, revealed that the dynamical law of the ejecta that matches free expansion (linear velocity law, De Gennaro Aquino et al. 2014) and not a wind (Friedjung 1966, 1987, 1992, 2011, Cassatella et al. 2004). 

Similarly, the HR spectroscopic monitoring of the transient ASASSN-17hx across its multiple light curve maxima (Mason et al. 2020) showed that the observed lines variations (both in profile and displayed ionic species) are consistent with a freely expanding ejecta subject to the irradiation from a varying underneath source. In particular,  
the ionizing source was harder during minima causing the ejecta pseudo-photosphere to recede, an increased excitation of the high ionization potential energy elements such as He, and the ionization of the low ionization potential metals (e.g. Fe~II and iron group).

A final example of the greater amount of information produced by the monitoring in HR spectroscopy is given by V1309 Sco (Mason and Shore 2022). In that case, the six UVES epochs allowed distinguishing two physically and dynamically distinct components, despite the low velocities of the individual spectral features. One, the low velocity component, was ascribed to circumstellar material, in view of the lack of evolution for over 345 days.  These spectra also constrained its geometry either to a gas region orthogonal to the expanding ejecta (non spherically symmetric) or to a very distant CSM with respect to the progenitor. Target of Opportunity (ToO) observations in Rapid Response Mode (RRM), could have helped constraining the distance of the CSM. The other component, that with high velocity, was identified with expanding matter based on its temporal development. In addition, a constraint could be set on the ionizing source which turned off before the ejecta became optically thin\footnote{The late monitoring was complemented with intermediate resolution spectra (XShooter), due to the faintness reached by the transient.}.

We stress, at last, that a single spectrum/epoch might be misleading also when it comes to classification since intrinsic peculiarities might apparently differentiate a transient from a given class identified through phenomenology. An example of this was ASASSN-18fv: initially suspected to be an exotic transient or a red nova (i.e. a non compact stellar merger, e.g. Strader et al. 2018, Izzo et al. 2018a), it turned out to be a nova once it entered the optically thin phase (UVES archive public spectra, see also Pavana et al. 2020). 
The plethora of very narrow absorption lines displayed by ASASSN-18fv made it strikingly similar to the discovery spectrum of V1309 Sco. Its bluer spectral energy distribution was not a particular concern in view of the fact that many extragalactic red novae are initially blue (e.g. Valerin et al. 2025a,b). Within a few months, ASASSN-18fv developed a spectrum rich in Fe~II lines with somewhat "castellated" profile, as typically observed during the optically thin phase of classical novae, and eventually displayed nebular transitions such as [Ne~III] and [O~III], also with similar profiles and no evidence of molecular bands, consistent with a passive ejecta ionized by a strong central source, as in novae. Compared to red-novae, this evolution suggests a smaller column density and different powering source. 

Monitoring of extragalactic rapidly fading transients with HR spectroscopy might not be feasible in short exposure times, but moving from HR to LR as the transient fades cannot help in this sense. It is, therefore, important to consider longer exposures (faint QSO and Ly$\alpha$ forests are routinely observed in HR spectroscopy) for at least a number of transients and transient types. 

\section{The limits of a small wavelength coverage}

Some astrophysical science cases are only marginally affected by a limited spectroscopic wavelength range but, for stellar transients requiring a physical characterization, it is important to cover the bulk of the transient spectral energy distribution (panchromatism). 

As mentioned earlier, the evidence for electron scattering requires the comparison of several emission lines, ideally from a same ion (e.g. H), to avoid biasing the inference with possible emission measure and/or recombination effects. Fig.\ref{f9b}, presented in Section~\ref{uves_a}, shows how the emission lines wings broadened by electron scattering affect each profile identically, producing same wings and velocity extensions\footnote{More precisely, it must be that the scattering and the line forming regions coincide in space or are one in front of the other.}. The S/N must be high across the whole wavelength range of the spectrum. 

Similarly, the line identification process discussed in Section~\ref{lrhr} ideally requires that each transition is confirmed by the simultaneous detection of other transitions from the same multiplets, or similar multiplets from the same ion, while recognizing that, in a (violently) expanding gas, LTE conditions do not apply and peculiar fluorescence effects (from lines or continuum) can occur that are induced by the velocity field of the expanding gas  (e.g., Williams \& Ferguson 1983, Shore 2012, Shore et al. 2014).  
Again, it is impossible to assess the ionization degree of a gas region from a small spectral range spectrum, since no species or ion has uniformly distributed transitions in any wavelength interval. 

The large majority of grisms (unless of extremely low resolution, R$\sim$100) cover only a few thousands \AA \ of spectral range, the blue or the red optical range, or H$\alpha$ and H$\beta$ and anything in between. But the blue (B) band alone, for example, cannot help assessing whether the detection of, e.g., N~III is due to fluorescence or recombination, or establishing, upon the detection of He~II, whether the ionizing radiation is limited to $\sim$54 eV or even higher, or correctly infer the density and/or temperature gradients. The red part alone, missing the blue wavelengths, cannot inform about metal absorptions (e.g. the iron curtain) or fluorescence effects (Bowen blend, O~III, O~IV Raman).  It further prevents the detection of most of the forbidden transitions used as density or temperature diagnostic (e.g. [O~II] at 3726,3729 \AA, [O~III]~5007,4959 and 4363 \AA, [Ne~III]~3869,3967 and 3342 \AA, [Ne~V]~3425,3345 and 2975 \AA, [S~II]~4068,4076 \AA). Detecting just the Ca~II NIR triplet does not establish whether it is due to recombination or the large opacity of the Ca~II H\&K resonance lines. It is interesting to note that many HR spectrographs, having exoplanet studies as the driver science case, tend to penalize the B and U band.  This  substantially impedes stellar science, in general, and the studies of transients, in particular. 

\section{Discussion and conclusions}
Taking advantage of the serendipity that the ESO archive contained almost simultaneous observations of the transient NGC~300 OT2008-1 taken at the VLT with the FORS and UVES spectrographs, we had the idea of comparing the two intrinsically identical spectra to show the different information that is possible to extract from each data set. For this reason, the two spectra have been analyzed independently and without accounting for the results already present in the literature to which we refer the reader interested in a comprehensive view about the object and its evolution (e.g. Berger et al. 2009, Kashi et al. 2010, Patat et al. 2010, Humphreys et al. 2011, Valerin et al. 2025a,b). 

Our analysis has two strong implications for studying transients. On the one hand it demonstrates the superior amount of information available in the HR spectrum with respect to the LR one. On the other hand, the reanalysis of the UVES spectrum of 300OT has revealed new insights not fully noticed before. The identification of the broad emission line region with a shock front and of the narrow absorption with circumbinary material produced by a companion star. In this new proposed scenario the outbursting object resides in a binary systems in which the companion has created a high column density environment. Because of the relative velocity of the shock emission and the narrow absorption from the undisturbed gas, the binary was viewed from one of the quadrant bracketing the 0.25 orbital phase at the time of the observations. 

This picture could not have been formulated based on LR spectroscopy. This is the great limitation of LR spectroscopy (and filter photometry): it impedes deductive analysis, favoring an overly simplified modeling based on (frequently, nonphysical) parameters fitting. 
In contrast, HR spectroscopy favors deductive analysis, securing the line identification with high accuracy, permitting accurate determination of the transient's ionization structure, allowing the identification multiple components (typically unresolved at low resolution, even in the case of $>1000$ km/s lines such as in novae\footnote{Until the early 2000s, novae were routinely observed in LR spectroscopy because their lines where resolved. However, it is the detection of their substructures and their kinematics that allowed determining the dynamics of nova ejecta and it is only through HR spectra that is possible to determine the ejecta filling factor.}) by pinpointing the true line profiles and wing extensions (often in LR spectra Gaussian or Lorenzian wings results from the instrument's LSF), and helping to constrain the gas kinematic and physical conditions\footnote{The ability to determine parameters such as the gas density or temperature depends upon optically thin conditions and the detection of diagnostic lines.}. However, determining the gas dynamics requires monitoring, i.e. a series of HR spectra sampling different evolutionary phases of the transient. Based on experience and the literature (e.g. Della Valle et al. 2002, Ederoclite et al. 2006, Williams et al. 2008, Mason et al. 2010, Shore et al. 2016), six to seven epochs per transient are sufficient to track its evolution; the cadence of the observations mildly depending on the transient's rate of decline in brightness. A S/N $>$20-30 along the whole spectrum is needed to sample line wings and weak features. 

The detailed HR spectroscopic followup of at least some individual transients requires a relatively large amount of time (of the order of tens of hours per object, in place of the usual few), and therefore a change of paradigm in the typical transient observing strategy. We are aware that the proposed approach is demanding in terms of (large) telescope time and oversubscription, but the gain is pivotal. In addition, those high quality data would serve as benchmark for sparse observations routinely collected for large sample of transients, especially now that the medium resolution, MR, spectrograph SoXS is will be dedicated to monitor the transient universe (Schipani et al. 2018). We are not claiming that HR spectroscopy will answer all the questions (in this sense HR spectropolarimetry provides even more information), but it is the key for a physical characterization of new or poorly known astrophysical phenomena. HR is hardly the solution for statistical and population analysis and in this sense different observing strategies are complementary. The transient community should, then, consider observations with large and extremely large telescopes and HR spectrographs not just exceptionally but somewhat systematically. 

When offered to the worldwide community, FEROS at the ESO 2.2m or FIES on the NOT,  were limited to $\sim$13 mag which made them suitable  for only bright Galactic transients  such as novae near maximum (e.g. Ederoclite et al. 2006, Williams et al 2008, Shore et al. 2011a, 2011b, 2012a, 2013, 2016, Mason \& Munari 2014). UVES on the VLT has been excellent for novae (e.g Mason \& Munari 2014, Molaro et al. 2016, Mason et al. 2018, Izzo et al, 2018b, Molaro et al. 2020) and other galactic transients during early decline (e.g. red novae Mason et al. 2010, Tylenda et al. 2011, 2015, Mason \& Shore 2022) but is limited to magnitudes around $\sim$17-18.  This clearly hampers the followup of novae at late decline (nebular phase) and the majority of extragalactic transients. Only the near maximum phases of bright Local Group transients can be observed with UVES. 

The ArmazoNes high Dispersion Echelle Spectrograph (ANDES, Marconi et al. 2024) is the second-generation, high-resolution, multi-wavelength spectrograph currently planned for the ESO Extremely Large Telescope (ELT). Conceived as one of the ELT’s flagship instruments, ANDES is designed to exploit the telescope’s collecting area by delivering ultra-precise spectroscopy across a very broad wavelength range, from the near-ultraviolet to the near-infrared (0.35 to 2.4 $\mu$m). With a baseline resolving power of R$\simeq$100000 and a fiber-fed design developed upon the heritage of instruments such as HARPS (ESO/La Silla) and ESPRESSO (ESO/VLT), ANDES is specifically optimized for exoplanet science, including the detection and characterization of exoplanet atmospheres and high-fidelity radial-velocity studies (Palle et al. 2025). However, this does not preclude ANDES from being a powerful instrument for a broad range of other astrophysical applications, including studies of galaxy formation and evolution, the intergalactic medium, stellar chemical archaeology, and time-domain astronomy, with particular emphasis on high-resolution investigations of transient phenomena (Roederer et al. 2024).

In its current design, however, the limiting magnitude is expected to be approximately 18–19, that is, only about one magnitude deeper than UVES. This limitation could constrain several of the non-exoplanet science cases, particularly those requiring the follow-up of faint or rapidly fading sources. For this reason, the ANDES consortium is actively evaluating possible strategies to improve sensitivity. One option under consideration is the deployment of an additional bundle of fibers connecting the front-end injection system to the spectrograph modules of ANDES, without the light scramblers that are necessary to limit systematics in precise radial-velocity measurements. The use of such fibers, without junctions, and revised spectral formats compatible with detector binning larger than 2×2, could enable observations of targets roughly one magnitude fainter (Schmidt, private communication), corresponding to an improvement of about two magnitudes relative to UVES. Although this may appear to be a modest gain, it would increase by more than a factor of fifteen the accessible volume of parameter space currently reachable with UVES. More importantly, it would extend the temporal baseline over which selected transient sources can be monitored, enabling observations into intrinsically fainter evolutionary phases and providing valuable insights on their physical and chemical evolution.

We conclude that HR spectroscopy, panchromatism (ideally UV+optical+NIR, or UV+optical and optical+NIR for blue and reddened transients, respectively), and good signal-to-noise ratio ($>$20-30 in the continuum) must be considered together. None of them, alone, can produce solid results. Nor can a single multiwavelength HR high S/N spectrum. 
A final extension, to HR linear spectropolarimetry, would be invaluable in constraining the structure of these complex sources.
The systematic use of those observing strategies, complemented by routine observations in low and medium resolution spectroscopy and photometry at smaller facilities, would provide  considerable improvements in the characterization of transients by combining the insights gained through a detailed physical analysis and those derived from statistical studies. It would, in addition, produce constraints for theoretical models, possibly reducing the forcing of oversimplified models to the observations.   

\begin{acknowledgements}
      EM thanks Ferdinando Patat from ESO-Garching for his precious suggestions about the data reduction of FORS linear spectropolarimetry data. EM thanks Tobias Schmidt for sharing his work about possible strategies to observe faint targets with ANDES. AP acknowledges support from the PRIN-INAF 2022 “Shedding light on the nature of gap transients: from the observations to the models”.
\end{acknowledgements}

%
%

\end{document}